\documentclass[acmlarge]{acmart}

\AtBeginDocument{%
  \providecommand\BibTeX{{%
    \normalfont B\kern-0.5em{\scshape i\kern-0.25em b}\kern-0.8em\TeX}}}

\setcopyright{acmcopyright}
\copyrightyear{2018}
\acmYear{2018}
\acmDOI{10.1145/1122445.1122456}

\acmJournal{POMACS}
\acmVolume{37}
\acmNumber{4}
\acmArticle{111}
\acmMonth{8}

\begin{document}

\title{In Alexa, We Trust. Or Do We? : An analysis of People's Perception of Privacy Policies}


\author{Sanjana Gautam}

\renewcommand{\shortauthors}{Gautam}

\begin{abstract}
 Smart home devices have found their way through people's homes as well as hearts. One such smart device is Amazon Alexa. Amazon Alexa is a voice-controlled application that is rapidly gaining popularity. Alexa was primarily used for checking weather forecasts, playing music, and controlling other devices \cite{garg2020he}.This paper tries to explore the extent to which people are aware of the privacy policies pertaining to the Amazon Alexa devices. We have evaluated behavioral change towards their interactions with the device post being aware of the adverse implications. Resulting knowledge will give researchers new avenues of research and interaction designers new insights into improving their systems.
\end{abstract}



\maketitle

\section{Introduction}
The Amazon Alexa is a voice-controlled application developed by the Amazon company for its Echo, Echo Dot and recently introduced Echo Show devices (Amazon, n.d.). Being marketed as an intelligent personal assistant (IPA), Echo/Alexa is reportedly used for playing music, answering general questions, setting alarms and timers, or controlling networked devices \cite{bentley2018understanding}. In the past two years, there has been growing interest in intelligent personal assistants (IPA) like Alexa and Google Home. Naturally, there has been some deliberation about how privacy plays a crucial role in affecting the usage of such IPAs. In the study \cite{lau2018alexa}, we see that privacy implications do not bother most of the users as they are satisfied with the convenience it brings to their life. Majority of these people are unaware of the privacy settings that Alexa is equipped with \cite{lau2018alexa}. We believe that users lack of concern stems from the ignorance about what can go wrong. We hope that the findings of this paper show us that educating the people about the repercussions has a very different effect on people compared to informing them of the rules. Any reinforcing event which serves as a reward for a learner has both motivational and informational functions \cite{estes1972reinforcement}. Efforts by representatives of society to modify people's behavior in socially desirable directions are all too often conducted in accord with the obsolescent view that rewards directly strengthen actions and punishments weaken them. Thus, in this study we make the participants aware of the potential punishments of their ignorance and rewards from cautious actions. One of the examples of the cases presented is, \textit{One interesting and amusing case, is the case of the six-year old girl from Dallas who prompted Alexa to order—much to her parents’ surprise—a \$160 Kid- Kraft Sparkle Mansion and four pounds of sugar \cite{shulevitz2018alexa}.} While this may be an innocent isolated case, there are cases of much more critical nature that are discussed in \cite{cox2019alexa} and \cite{esq27alexa}. In \cite{cox2019alexa}, we see that the query history of Alexa was used to solve a murder trial. Accessing the phone records of the accused led the police to his Alexa search history. Going forward we will see examples that show the implications of privacy breach go beyond using the information for commercial reasons. The information about heat and light sensors linked to the device in the wrong hands can have disastrous effects. The following question would then be why focus only on Alexa in this study? With approximately 8.2 million Echo family devices sold since 2014, Amazon controls 70\% of the intelligent personal assistant market. Amazon’s Alexa Voice Service (AVS) provides voice control services for Amazon’s Echo product line and various home automation devices such as thermostats and security cameras \cite{ford2019alexa}. Access to personal home devices like thermostats and security cameras makes Alexa a critical tool in the hands of the wrong person. 
\\
The goal of the project is to create awareness among users about the tools available to their benefit and also to understand how their privacy choices change when presented with new facts. In this paper, we further explore what is the immediate response of the users after they become aware of the privacy implications. This could give us an indication of what steps to take to make sure the users are aware and attentive about their information privacy. Towards the end we discuss design recommendations for the IPA devices that can aid the awareness process. This can help both the user and the company build a trusting relationship. 

\section{Literature Review}
As we move forward with the paper, we realize that there are many facets to this problem.  We divide our literature study into three sections. The first section summarizes the finding from all the papers about how people interact with Alexa. It further talks about how much Alexa has penetrated into the regular lives. The second section narrates how the privacy implications go beyond just commercialization, for example, how cases have been resolved using the data from Alexa. The last section summarizes the previous studies that revolve around the security and privacy issues pertaining to Alexa. Our work further builds on the past studies. 

\subsection{What is Alexa to People ?}
Home life has a daily rhythm from waking up in the morning, leaving for the day, coming home, entertaining, and going to sleep. As conversational agents intersect with these different moments, users interact differently with them \cite{sciuto2018hey}. The placement of the device in home has been subject to uncertainty. There are studies that clearly indicate bedroom emerges as the most popular placement choice. At the same time, there are other studies that suggest that kitchen and living room may be more preferred \cite{lopatovska2019talk}. Another interesting aspect is how children at home respond to the device. It was seen in \cite{sciuto2018hey} that very young children are highly influenced by the spoken nature of conversational agents, respond to the agents at a very young age, and imbue the agent with human-like qualities. It then becomes important to understand how Alexa works. Alexa’s aptitude for voice recognition is improved by the collection of data. Built with artificial intelligence capable of natural language processing, recording questions and other requests made by the Amazon Echo user help Alexa to develop a better understanding of language nuances such as accents and dialects \cite{orr2018alexa}. Given that the Amazon Echo is powered by the cloud-based Alexa Voice Service, all collected details stored on Amazon servers which may be located outside of the country in which a user resides. Things brings to focus data protection laws cross borders. However, it isn’t in the scope of our current work. The Amazon Echo has several affordances that predispose users to personifying it and integrating it into social life. First, to operate the device, users must interact with it \cite{purington2017alexa}. Upon analysing the top reasons why people indulge in Alexa, entertainment emerged as the clear winner. It was followed by information seeking and daily assistance (calendar and alarms), while companionship was the least frequent. Thus, the Echo is inherently a socially interactive device \cite{fong2003survey}. Even after training on massive data, it really is remarkable how often smart speakers flub their lines. They’ll misconstrue a question, stress the wrong syllable, offer a bizarre answer, apologize for not yet knowing some highly knowable fact. Alexa’s bloopers float around the internet like clips from an absurdist \cite{shulevitz2018alexa}, yet they remain extensively popular. 

\subsection{Legal Defence among Privacy Concerns}
Although Alexa provides convenience to its users, it undermines and diminishes one’s privacy through its third-party inter-connectivity. In order to get full use out of Alexa’s skills, the user will ask Alexa’s queries or commands \cite{lau2018alexa1}. Under the third-party doctrine, an individual does not have a reasonable expectation of privacy regarding the information that he or she voluntarily disclosed to a third-party \cite{lau2018alexa1}. Though there are rules and regulations in place to prevent misuse of the information by these third-parties. Fourth Amendment evolved through case law to provide citizens with protections when in the intimacy of one’s home \cite{lau2018alexa1}. Despite these protections, the third-party doctrine peels away a person’s reasonable expectation of privacy when data or information is exposed to third parties  \cite{cox2019alexa}.  Evident by the legal concerns arising from cases like Arkansas v. Bates, and State v. Verrill  \cite{cox2019alexa}, it is critical that  our  laws  evolve  in  lockstep  with  technology  to  maintain the protections our Constitution provides citizens. While the Fourth Amendment does provide protections, it is only against unreasonable searches and seizures. Therefore, an intrusion may not violate the Constitution if the government first obtains a warrant with probable cause pursuant to the Warrant Clause of the Fourth Amendment. In contrast, when information leaves the home and is shared with third parties, courts have held that there is no reasonable expectation of
privacy \cite{pfeifle2018alexa}. Thus, the advancement of technology makes applying the third party doctrine difficult because more people are entrusting their communications to third parties. Alexa is constantly in listening mode, waiting for the "wake word" to be spoken. You configure Alexa to respond to one of the four wake words ("Alexa," "Amazon," "Echo" or "computer") \cite{nelson2017alexa}. The always-on microphone has presented the biggest privacy issues for voice assistant enabled devices. Several incidents involving privacy breaches have already occurred: Google Home Mini was found to be recording when the wake up word had not been used; Amazon's Alexa recorded a family's conversation and sent it to a contact; and both Alexa and Google have been triggered by either radio or television commercials\cite{cox2019alexa}\cite{shackleton2018alexa}. 

\subsection{Privacy Perceptions of Smart Speakers}
Prior research has indicated that microphones are perceived as one of the most intrusive sensors in home contexts \cite{bugeja2016privacy}. This juxtaposed against the observations from studies that indicate vast majorities unaware of privacy controls offered by the device \cite{lau2018alexa}. Some personal factors that aid our decision making with regards to online privacy are : trust in the provider, degree of enjoyment derived from the platform, trust in the other members of the platform and privacy awareness \cite{krasnova2012self}. The reasons why people trust Alexa may go deeper into the fundamentals of how human trust works. Trust may not always operate but may be more likely to operate under extreme conditions, where very sensitive data or, conversely, innocuous data are being requested from very well-known companies or, conversely, companies that are easily identified as “questionable’ \cite{norberg2007privacy}.  In a study of users and non-users,  \cite{norberg2007privacy} users adopted smart speakers for the convenience they would provide or because they liked to be early adopters. Non-users either did not find enough utility in the devices to warrant adoption or had privacy and security concerns regarding smart speaker use. Some people reason their lack of privacy concern to the fact that the amount of resources required to store their day to day information is too expensive and outweighs the benefit to the company. It further attributes the misguided use of the device to retroactive nature of controls than the otherwise preferred proactive control. Users express few privacy concerns, but their rationalizations indicate an incomplete understanding of privacy risks, a complicated trust relationship with speaker companies, and a reliance on the socio-technical context in which smart speakers reside \cite{lau2018alexa1}. Current smart speaker privacy controls are rarely used, as they are not well-aligned with users’ needs. Even if companion apps use encrypted network connections, sniffing the traffic between the apps and the IVA can expose the ecosystem’s communication mechanisms. Better diagnostic testing of IVA ecosystems can reveal vulnerabilities and lead to more trustworthy systems \cite{chung2017alexa}. While there are applications for mobile devices that can help personalize privacy experience \cite{liu2016follow}, there are no such technologies for smart home speakers. The said application shows promise and effectiveness. It may be possible to extrapolate design solutions from it.

We see that the papers have covered the lifestyle related analysis of Amazon Alexa. We also see a lot of precedence related literature discussing what kind of laws exist to defend the users. And towards the end we see how people have very varied perceptions about how Amazon protects their privacy. However, none of the papers look to analyze the nudges that can get the users thinking about their privacy and the steps they can take to protect it. In this paper, we evaluate the awareness level of the participants. We also look to analyze if the participants are willing to alter their behavior if provided with more information concerning the privacy policies. 

\section{Research Objective}
This work brings together the findings from the preliminary studies that indicate people’s lack of awareness about Amazon Alexa’s privacy policies and further studies any adaptive change in behaviour.  We explore the extent to which people are aware of the privacy policies. It also tries evaluate any behavioural change towards their interactions with the device post being aware of the implications. The research question we are looking to address is, would complete awareness of the potential privacy risks of IPAs (Intelligent Personal Assistants) affect the way we use them?

\section{Methods}
We believe that the past work has highlighted the lack of information about the way information and personal interactions with IPAs are stored and managed \cite{nelson2017alexa}\cite{orr2018alexa}. This has had an impact on the studies previously conducted like \cite{lau2018alexa}. We expect that there would be a significant difference in the way people approach these everyday technologies once they are aware. This work could potentially contribute in getting feedback on methods to improvise the existing privacy settings. We can extrapolate the findings on the relation between awareness and behavior to other AI based technology.

We have used the mixed methods approach. Data collection would be done leveraging the two tools : survey and interview. Survey is chosen to reach out to as many participants as possible. Considering the short span of the study, surveys can cover more grounds at a faster rate. This was a cross-sectional survey i.e., the data would only be collected at one point of time.  For the purpose of survey, Qualtrics was used to design as well as distribute the survey. The interview followed a semi-structured approach. Same set of questions would be addressed to all the participants. However, based on individual responses further questions emerged. All participants are required to be above 18 years of age and should own one of the Amazon Alexa device (i.e., Echo Dot, Echo Show, Echo). The study has been cleared by IRB to come under exempt status.

\subsection{Participants}
 We recruited participants by invitations and advertising on Facebook and other social media, and personal connections. For the interview, there were three male and two female participants. Their age group varied from 18 years to 51 years. All the participants were required to be fluent in English as the interview was conducted in English. Participant consent was taken prior to recording the interview audio. 

\subsection{Quantitative Analysis Approach}
The survey is divided into five sections. The first sections probes the basics of their usage patterns. The second section probes to their habits with regard to privacy policies. The third section describes the scenarios that paint a vivid picture of the privacy implications. This section is specially added because previous studies indicated lack of understanding of privacy concerns. The fourth sections re-evaluates the stand of participants on privacy controls. The last section evaluates their knowledge about privacy controls. For quantitative analysis, we captured the shift in decisions made by the participants after being introduced to the new information. If it is significantly high, our hypothesis is validated. We use within-subjects design so that the variance due to participants’ predispositions will be approximately the same across test conditions. Apart from collecting the demographic details, we want to understand their usage patterns and privacy concerns. We want to capture their purpose of use and also how dispensable or indispensable it is to their lives. This would give us an indication of how much they value the utility of the product over the privacy cost. 

Described  below are the scenarios that were used as a part of the survey. They were curated after a thorough study of the literature to identify the cases of all magnitude ranging from harmless to menacing. These specific scenarios were chosen after running a pilot study with five participants. It was observed that increasing the number of case studies to more than four resulted in loss of attention. Thus, we limited the cases enlisted. All the scenarios are true i.e., they are real-life incidents reported in past literature or newspaper articles. We anonymized the details to maintain the privacy. Special care was taken to describe the scenarios to get a complete understanding of the situation while not making it verbose. 

Scenario 1: \textit{A man convicted of first-degree murder, inter alia, is an exemplar of how invasive and incriminating smart devices can be. As law enforcement began interviewing the man, they asked for his consent to search his phone. The phone extraction revealed that the man had a Google application on his phone. The extraction also provided the password to his Google account. The detectives then logged into his account from a separate computer, which rendered access to his account. The detectives listened to his automated recordings of his google searches and were able to identify his voice. The searches revealed evidence that insinuated he committed the murder.}

Scenario 2: \textit{When a person using Amazon’s voice assistant in Germany requested to listen to his archive of recordings, he got much more than he was expecting. Along with receiving his own audio history captured by a home microphone, the user also gained access to 1,700 audio files from a person he did not know. Amazon sent the man a link that contained a stranger’s recordings, allowing him to listen to another man speaking inside his home with a female companion. }[8]

Scenario 3: \textit {Your smart home device could possibly has access to your smart lights, smart TV, thermostat and so on. Your smart home device can be activated by voice controls. Anyone loud enough to reach out to the device can control your home appliances. This means that if your device is placed in a common area, your neighbor can potentially control these devices remotely.  }

Scenario 4: \textit{A 6-year-old Dallas girl sharing her love of dollhouses and cookies with the family’s new Amazon Echo Dot prompted Alexa to order—much to her parents’ surprise—a \$160 Kid-Kraft Sparkle Mansion and four pounds of sugar cookies. After reporting the story, the anchor of a San Diego TV morning show remarked, “I love the little girl saying ‘Alexa ordered me a dollhouse.’” Several Echo owners watching the broadcast reported that, after hearing the anchor’s comment, their own devices also tried to order pricey dollhouses.}

\subsection{Qualitative Analysis Approach}
As a significant part of the work is qualitative, we are looking to identify the themes that define the ‘Trust in Alexa’ among users. The interview acted as a follow-up to the survey. The interview questions probed into the multi-user scenarios, their mental model of smart speakers, reasons for adoption and privacy concerns. All interviews were transcribed using a transcription service, with subsequent quality control by the authors. 

\section{Analysis and Results}

\subsection{Survey Analysis}
There were total 64 responses collected. The data was then cleaned to eliminate any incomplete and misreported surveys filled. After all the filtering we were left with 57 survey responses. The survey can be divided into five sections with a total of 28 questions and 4 scenarios. In the first section, we explore details pertaining to the demographic and usage pattern. The following section probes their current understanding of how Alexa processes and stores their personal data. We then introduce the section that is used for pre and post analysis. This section has questions centering around the theme of the trust in the device. The next section presents to the participants the four scenarios followed by the post-analysis questionnaire. At the end, we also inform the users about the basic data privacy policies that the company follows. Moving forward we analyze the survey responses section by section.
\\
\begin{figure}[h]
\caption{ Regular Users in Individual Households}
\includegraphics[scale=0.65]{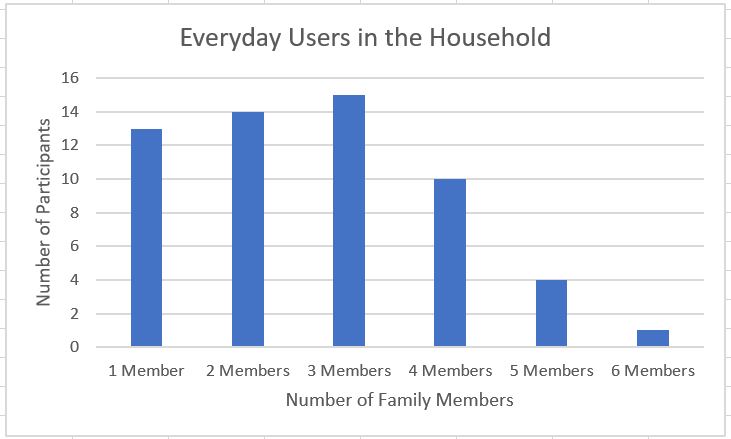}
\centering
\end{figure}
\\
As discussed above, the first section the usage pattern among users is extracted. Some of the information is represented in graphical form. Most of the households reported 2 - 3 regular users (refer Figure 1) using it 2 - 5 times a day (refer Figure 2). Alexa as a source 'Entertainment' was the most common use of the device. Some participants specially mentioned the user of Alexa device particularly for 'listening to music' and 'weather reports'. It was also found that 24\% of the participants had a smart home appliances linked to their PVAs. Although only 1 one of the participant had a child below 10 years of age who had regular access to the PVA device. 
\\
\begin{figure}[h]
\caption{Frequency of Usage in the Household}
\includegraphics[scale=0.65]{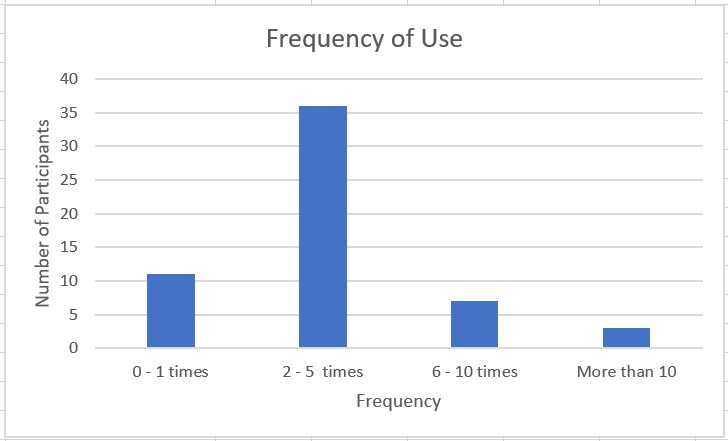}
\centering
\end{figure}
\\
Participants were asked to self-report their awareness on the Privacy FAQs. It was seen that 38\% of the participants were 'Definitely Not' aware of them, followed by another 26\% 'Probably Not'. This brings the total tally of unaware participants to 64\%. Only 14\% of the participants could say with confidence that they were aware of the privacy related FAQs. When we asked the participants if they think their PVAs record all of their conversations 60\% of them were inclined towards believing that it does. This number changed to 78\%  when the conversations that were followed by the 'wake word' were discussed. 
\\
\begin{figure}[h]
\caption{ Self-Reported Awareness of Privacy Settings }
\includegraphics[scale=0.65]{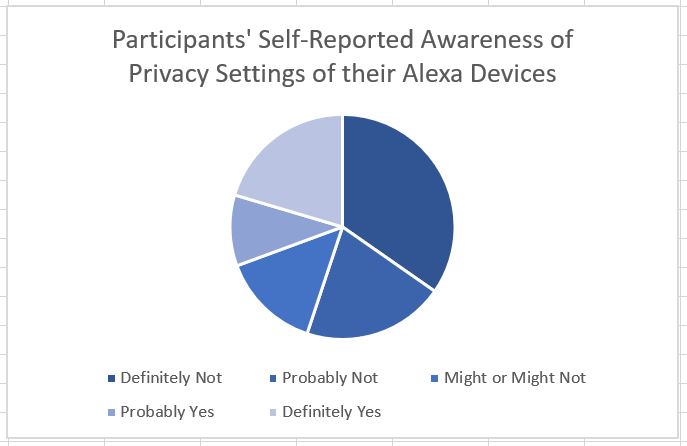}
\centering
\end{figure}
\\
36\% of the respondents were unaware that they could mute their PVAs and 82\% of the participants were unaware the duration for which the conversations are stored in the cloud. We also asked the participants about the access permissions that the device has that is known to them. It was seen that an average of 75.92\% of the participants were aware of the specifics. 
\\
Now, we discuss the questions that inquire into the feelings of safety and trust among the participants. We performed t-Test analysis on the responses by dividing the samples as before scenario and after scenario. Table 1 describes the results from the T-Test. We see that people were encouraged to find out more about the privacy settings of their PVAs after reading the scenarios. The initial percentage of 30\% jumped to 91.7\%. However, the same enthusiasm was not shared when it came to reading the 'Terms and Conditions' that came with the device. It jumped from 16\% to 47.9\%. 

\begin{table}[]
\centering
\caption{t-Test Results for Trust and Safety}
\begin{tabular}{|l|l|l|l|l|}
\hline
                 & \multicolumn{2}{l|}{Trust}        & \multicolumn{2}{l|}{Safety}       \\ \hline
                 & Pre              & Post           & Pre              & Post           \\ \hline
Mean             & 4.05             & 3.37           & 4.49             & 3.67           \\ \hline
Median           & 4                & 3              & 5                & 3              \\ \hline
p-value          & \multicolumn{2}{l|}{0.00162}      & \multicolumn{2}{l|}{0.0000975}    \\ \hline
CI of Difference & \multicolumn{2}{l|}{0.27 to 1.09} & \multicolumn{2}{l|}{0.43 to 1.20} \\ \hline
\end{tabular}
\end{table}

\begin{table}[]
\centering
\caption{t-Test Results for Comfort Around the Device}
\begin{tabular}{|l|l|l|l|l|}
\hline
                 & \multicolumn{2}{l|}{\begin{tabular}[c]{@{}l@{}}Comfort with \\ Standby Mode\end{tabular}} & \multicolumn{2}{l|}{\begin{tabular}[c]{@{}l@{}}Trust with \\ Private \\ Conversations\end{tabular}} \\ \hline
                 & Pre                                         & Post                                        & Pre                                              & Post                                             \\ \hline
Mean             & 3.38                                        & 2.60                                        & 3.31                                             & 2.60                                             \\ \hline
Median           & 4                                           & 2                                           & 4                                                & 2                                                \\ \hline
p-value          & \multicolumn{2}{l|}{0.0000460}                                                            & \multicolumn{2}{l|}{0.00731}                                                                        \\ \hline
CI of Difference & \multicolumn{2}{l|}{0.43 to 1.12}                                                         & \multicolumn{2}{l|}{-1.22 to -0.20}                                                                 \\ \hline
\end{tabular}
\end{table}

We see that in all the cases described in the Table 1 \& 2 the t-test report p-values that are significant. The paired sample t-test, sometimes called the dependent sample t-test, is a statistical procedure used to determine whether the mean difference between two sets of observations is zero. In a paired sample t-test, each subject or entity is measured twice, resulting in pairs of observations. The paired sample t-test has four main assumptions: The dependent variable must be continuous (interval/ratio). The observations are independent of one another. The dependent variable should be approximately normally distributed. The dependent variable should not contain any outliers. Common applications of the paired sample t-test include case-control studies or repeated-measures designs. The H0 is there is no difference between the two groups while H1 is that before > after. We see the p-values reject H0. 

\subsection{Interview Analysis}
There are five interview studies to be analysed. We automatically transcribed all the interviews and manually verified them. The interview was designed to further reason out the results from the survey. The questions were designed to extract 'why' behind the answers obtained from the survey. One of the most crucial questions that if the case studies have had any impact in their behavior. A deductive approach was adopted based on the survey results. A semantic analysis was done on the transcribed interview results. The interview data was coded and the themes were identified. We discuss the various themes that have emerged from the interviews we conducted. 

\subsubsection{Admitted Unawareness about Alexa}
Having Alexa at home was also a cause of concern for some people, as we can deduct from Harvey's comments, '\textit{having an Alexa Echo in my own room brings certain questions come in mind, mostly I feel safe around it but it's not completely safe as it may be hearing all our conversations every moment.}' We see that misconceptions and partial information about the device creates fear in people. It was seen that people admitted not being aware of the privacy policies. However, in the case they were aware, they could get comfortable with the device. Anna recalls, '\textit{an year ago when my friend mentioned that Alexa is, like, listening to everything that I'm saying but I found out that they basically only record what you say after you say the word 'Alexa' so I'm actually pretty comfortable with the privacy policies.}'

\subsubsection{In the Brand they Trust}
It was seen that participants relayed their trust onto the brand more than the device mechanisms in particular. One of the participant, Louis, commented that '\textit{ I am actually don't trust the methodology or the mechanics of it (the device) I would rather go by the brand}'. The integrity and ethics of the company is important to the users in building the trusting relationship. They attributed the belief in brand because those brands already have a lot of information about them, Like Anna says,'\textit{ So eventually I know that Amazon, of course, knows what I do with my account when I buy what I what I say but I am I would like to think that they only use it for marketing and for you know Amazon related thing.}' Trust has been defined previously in a variety of ways \cite{norberg2007privacy}. In our case the trust is often defined by the brand. Amazon being a huge brand, is trusted by it's customers. 

\subsubsection{Simple Words Go Far}
All the participants accepted the lengthy nature of the 'Terms and Conditions' deterred them from gaining a full insight of what grounds it captures. All the participants agreed that they would prefer these documents be laid out in more coherent way sans the legal jargon. There is past work discussing in details the tools that have been used to summarize legal texts \cite{polsley2016casesummarizer}\cite{yang2008hierarchical}. They also believe that the policies were presented in such a convoluted manner so that it is not accessible. 

\subsubsection{Takeaways from the Study}
It was observed that most of the participants we interviewed were being careful about their habits when it came to PVAs. They were already taking the required care and precaution. It should also be noted that they had limited use from the device. However, one of them, Harvey mentions, \textit{most part of the day I would like to keep the mic off. And (switch it on) whenever I have to use it... I will just stick to that purpose.} 

\section{Discussion}
Our results show that there is significant immediate response to the information that the participants were exposed to. We also see that participants though had an instinct about the privacy concerns were not aware of the real extent of potential damage that the device can incur. The survey also proved to us that their initial level of trust was not very high. On a scale of one to seven the average on trust is 4.05 while on safety it was a little higher at 4.49. Both those scales dropped to a mere 3.37 and 3.67 average. When we asked them if they felt comfortable having private conversations around their IPA device. The average for that dropped from 3.31 to 2.60. For any product, the trust and safety of the customer in it is of huge importance. So, this work is a step in the right direction towards simplifying the process of accessing privacy related information. Now, from the interviews we got the sense that participants were negligent towards their privacy.There are past studies that have also pointed towards this. For example, in \cite{sundar2013unlocking} we saw that users in general are positively disposed towards personalizing systems and will enter personal information into them unless they are primed with heuristics related to specific privacy violations such as fuzzy boundary. This tendency for privacy compromising online behavior in the face of privacy concerns is known as “privacy paradox”. Real-time accessible cues can help the users by 'nudging' them towards the a more aware decision. Another one of the design suggestion could be visual prompts that indicate clearly that the device is active and recording and transcripts that get reported back to the customer. For a company like Amazon, the findings from this paper indicate that prompt cues can help the users make smart calls. We understand there is a trade-off between providing the excess information and required awareness. As a next step, studying the behavior in real-time would give us a chance to understand the participant activity surrounding the device. This could lead to more substantial design implications. At this point, we might not be aware if our stimulation have any actions. It requires further experimentation. 

\section{Limitations}
We used Facebook as our main methods of participant recruitment owing to the special circumstances with the spread of COVID-19. This restricts our participant sample to the social network that was reachable within my profile. The demographics are swayed mostly towards youth between the ages of 19 - 22. Also, survey fatigue might have resulted in some inaccurate results. The case studies were long and verbose. It is understandable that people would not have paid close attention to them. Ultimately, both the survey and the interview were self-reported. They carry the limitations that come along with self-report. 

\section{Future Works}
Going forward it would be useful to actually monitor people's behavior around their smart devices. This can provide us much more accurate report on any behavior changes that might stem from better awareness about the privacy policies. Although we saw an immediate change in their attitude towards home device privacy, we cannot at this point conclude a sustained interest. Going forward we can do a follow-up survey with the same participants to understand if it is a temporary or permanent change. Further research is also required when dealing with summarizing the 'Terms and Conditions' to a simple elegant and easy to grasp language to improve the reach of such policies. 

\section{Conclusion}
In a foregoing time, a person had control over their own information—personal documents of banking or business transactions existed on a sheet of paper that the person possessed in their file cabinets. The scribbles and intimate handwritten notations in one’s diary would be safely tucked away under their pillow or buried in their nightstand. But with changing times a lot of our information is already on the web via all the social networking sites. With devices penetrating our lives, it is within our rights to stop more of our information going out there. In this study, we have seen that there is substantial unawareness about privacy settings about their smart home IPAs. We also see that the participants are willing to alter their behavior upon being made aware. This work is the stepping stone towards coming up with design solutions that can go a long way in protecting private information. 


\bibliographystyle{ACM-Reference-Format}
\bibliography{sample-base}

\appendix

\end{document}